\documentclass[a4paper,11pt]{article}

\usepackage[T1, T2A]{fontenc}
\usepackage[utf8]{inputenc}
\usepackage[main = english,russian]{babel}
\usepackage{cmap}
\usepackage{graphicx}
\usepackage{booktabs}
\usepackage{jinstpub}
\usepackage{tabularx}
\usepackage{array}
\newcolumntype{Y}{>{\centering\arraybackslash}X}
\graphicspath{{figures/}}

\title{Neural-network-based material identification in photon-counting computed tomography using region-of-interest spectral features}

\author[a]{A.~Miadzvetski}
\author[b,c,*]{V.~Rozhkov}
\author[b,c]{R.~Sotenskii}
\author[c]{D.~Shashurin}
\author[b,c]{G.~Chelkov}

\affiliation[a]{International Sakharov Environmental Institute of Belarusian State University, Minsk, Belarus}

\affiliation[b]{International intergovernmental research organization Joint Institute for Nuclear Research, Dubna, Russia}

\affiliation[c]{Lomonosov Moscow State University, Moscow, Russia}

\emailAdd{rozhkov@jinr.ru}

\abstract{
This work investigates the feasibility of neural-network-based material identification based on photon-counting computed tomography (PCCT) data. The input data were spectral vectors extracted from regions of interest (ROIs) in reconstructed tomographic slices of a phantom containing La-, Nd-, and Gd-based samples, as well as air, water, bone, and polymethyl methacrylate (PMMA). The original tomographic slices were acquired at 12 detector threshold settings (THL = 45, 55, 65, 75, 85, 95, 105, 115, 125, 135, 145, 165). For model training, the spectra were interpolated onto a uniform grid from THL 45 to THL 165 with a step of one THL unit. Large ROIs, small ROIs, and their combined dataset were considered. To avoid an overestimated performance caused by correlated ROIs extracted from the same tomographic slice, the data were split into training, validation, and test subsets at the slice level.

The MLP and 1D-CNN architectures were first compared using the raw threshold-response representation. Several preprocessing strategies were subsequently evaluated for the MLP, including raw spectra, logarithmic transformation, area normalization, and a combined representation comprising the logarithmically transformed spectrum and its first derivative with respect to THL. An additional experiment evaluated \texttt{Empty} as an auxiliary background class. The best performance was achieved using an MLP trained on the combined set of large and small ROIs with the \texttt{log1p\_deriv} feature representation. For the 10-class material-classification task, the final model achieved an accuracy of 0.9745, a balanced accuracy of 0.9745, and a macro-averaged F1-score of 0.9745 on the test set. The highest F1-scores were obtained for \texttt{Air}, \texttt{Bone}, and La\_2. The main residual errors were associated with the H$_2$O/PMMA pair and with several pairs of La-, Nd-, and Gd-based samples. Including an \texttt{Empty} class reduced the macro F1-score to 0.9294, mainly due to confusion between \texttt{Air} and \texttt{Empty}; therefore, \texttt{Empty} was excluded from the main material-classification task and treated as an auxiliary background class.
}


\begin{document}
\maketitle
\flushbottom

\section{Introduction}
\label{sec:introduction}

Photon-counting computed tomography provides information not only on the spatial distribution of X-ray attenuation, but also on the spectral response of the imaged object. This method uses photon-counting detectors (PCDs) enabling energy discrimination directly at the detector level~\cite{taguchi2013vision,llopart2007timepix,ballabriga2013medipix3rx,dudak2020timepix,procz2013medipix3ct}. PCCT is being actively assessed in medical imaging, as well as in other areas requiring non-destructive methods of analysis and visualization.

One important application of PCCT is material identification, where different substances may have similar densities or similar attenuation values at individual energy thresholds while exhibiting distinct spectral responses. This is especially relevant for samples containing high-$Z$ elements with pronounced spectral features, including iodine, gadolinium, and other elements used as contrast agents~\cite{suslova2022nanomaterials,suslova2022lafml}. Early proof-of-concept studies demonstrated the feasibility of material-specific K-edge imaging with PCCT in controlled phantom experiments~\cite{schlomka2008kedge}. Subsequent studies extended this approach to in vivo preclinical imaging. In particular, gold K-edge imaging was used for the detection and quantitative assessment of macrophage-rich atherosclerotic plaques while separating gold nanoparticles from iodinated contrast material and calcifications~\cite{simohamed2021atherosclerosis}. Gadolinium K-edge angiography was also demonstrated in atherosclerotic rabbits, enabling selective visualization and quantitative assessment of small vessels after administration of a gadolinium-based contrast agent~\cite{boccalini2023gadolinium}. Recent research has also focused on the development and optimization of dedicated high-$Z$ contrast agents for K-edge PCCT, with potential applications in vascular, oncological, and molecular imaging~\cite{jost2023kedge}. These studies demonstrate the potential of K-edge PCCT not only for material separation under controlled phantom conditions, but also for quantitative vascular and molecular imaging in vivo. 

Existing approaches to material-sensitive processing of photon-counting CT data are commonly based on the selection of a limited number of energy windows, K-edge analysis, or material identification methods for a predefined set of basis components~\cite{clark2014spectraldiffusion}. Such mathematical methods are physically interpretable and can be effective for detecting a specific element or estimating its concentration. In particular, for a closely related experimental setup, an algorithm for qualitative and quantitative material analysis using K-edges was proposed, with the aim of detecting the contrast agent and estimating its concentration~\cite{Sotenskii_2024}. However, they require prior knowledge of the element of interest for the selection of appropriate spectral windows, as well as a detector-response model and a predefined set of basis materials. Moreover, when different samples have close spectral vectors, such methods may be insufficiently sensitive to weak differences related not only to elemental composition, but also to the matrix, solvent, density, or sample-preparation procedure.

Machine-learning-based approaches provide an alternative to conventional mathematical methods of material identification in PCCT. Such approaches have been actively developed in recent years~\cite{abascal2021material,bousse2024review,rajagopal2025multimaterial}. However, their capabilities and generalizability require further investigation. An advantage of machine-learning-based approaches is that they do not require the manual selection of a small number of energy windows representative of a specific element of interest and can instead use the full spectral pattern. Their main limitation is their dependence on the training dataset: the model solves a classification problem for predefined classes and is not a universal method for determining arbitrary elemental concentrations outside the domain represented by the training data. 

The aim of this work was to develop and evaluate a neural-network-based approach to material identification from PCCT data through multi-class classification of specific materials and samples, including spectrally similar sample pairs, using the full region-of-interest (ROI) spectrum. In this formulation, the model must distinguish not only different elements, but also different samples containing the same element, provided that their experimental spectral responses differ. For this purpose, spectral features were constructed from measurements acquired at different THL settings, several spectral preprocessing strategies were compared, and two neural-network architectures were trained and evaluated: a multilayer perceptron (MLP) and a one-dimensional convolutional neural network (1D-CNN). The MLP treats the ROI spectrum as a single feature vector and can account for global relationships between different THL values. The 1D-CNN treats the spectrum as a one-dimensional signal and can potentially extract local patterns in the shape of the spectral curve. 

The main tasks of this work were:
\begin{enumerate}
    \item to construct a reproducible set of spectral features extracted from large and small ROIs;
    \item to implement a data-splitting strategy that prevents correlated observations from the same tomographic slice from appearing simultaneously in the training and test sets;
    \item to compare MLP and 1D-CNN models under the same training and evaluation protocols;
    \item to evaluate the effect of spectral preprocessing on classification performance;
    \item to identify the main sources of residual classification errors, including close material pairs and samples containing the same target element.
\end{enumerate}

\begin{figure}[htbp]
    \centering
    \includegraphics[width=0.98\textwidth]{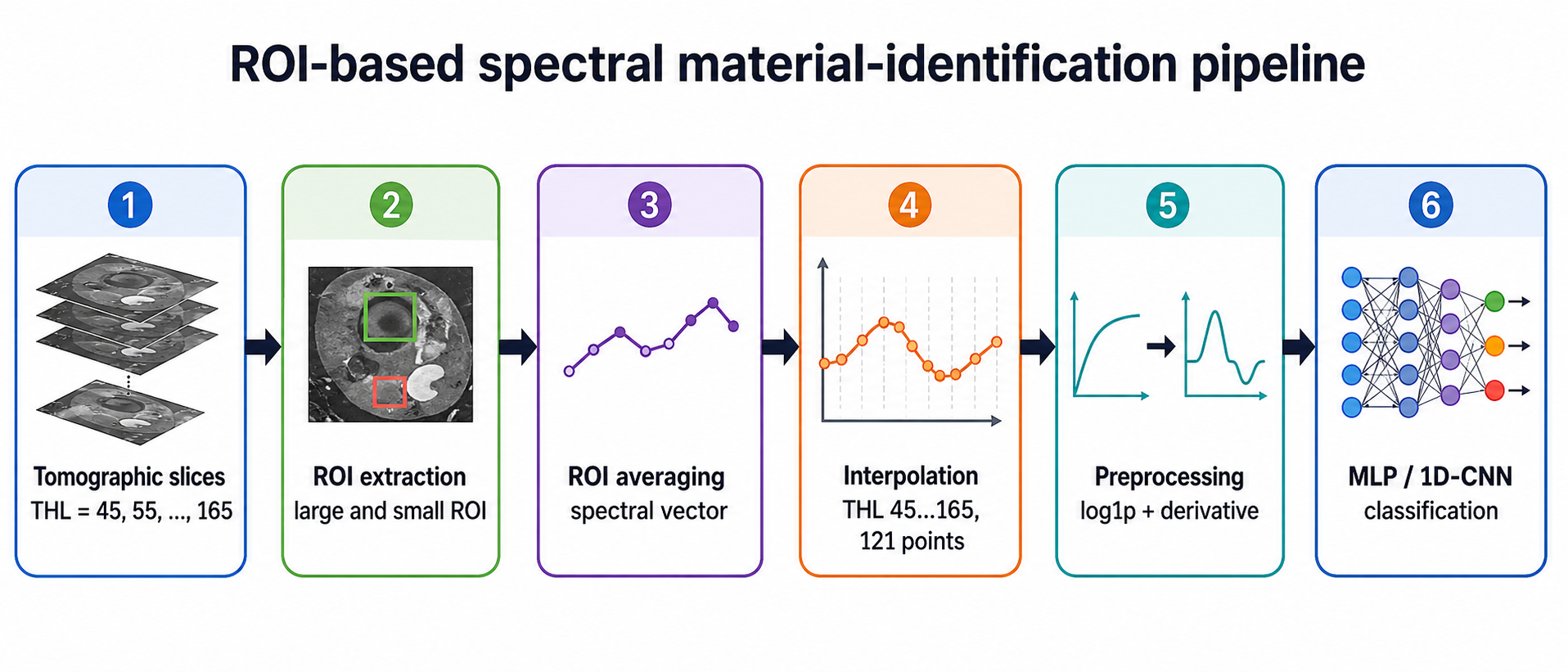}
    \caption{Workflow for forming spectral features and training the material-classification model. The input data consist of a series of tomographic slices acquired at THL of 45, 55, 65, 75, 85, 95, 105, 115, 125, 135, 145, and 165. After ROI extraction, the signal is averaged, the spectrum is interpolated, the features are preprocessed, and the neural-network model is trained.}
    \label{fig:pipeline}
\end{figure}

The general data-processing workflow is shown in figure~\ref{fig:pipeline}. It includes acquisition of a series of tomographic slices at different THL values, ROI extraction, formation of a spectral vector within each ROI, feature preprocessing, and subsequent training of the classification model.

\section{Materials and data}
\label{sec:materials}

\subsection{Investigated phantom}
\label{subsec:phantom}

The dataset used in this work was obtained from a phantom containing La-, Nd-, and Gd-based samples, as well as background materials. A closely related phantom configuration and a K-edge-based material-analysis algorithm for PCCT were previously described in~\cite{Sotenskii_2024}. The main classification task included 10 material classes:
\begin{itemize}
    \item \texttt{Air} --- air;
    \item \texttt{Bone} --- bone sample;
    \item Gd\_1 --- composite contrast agent consisting of Gd$_2$O$_3$ nanoparticles stabilized on a matrix of few-layer graphite fragments (FGF) with a Gd concentration of 80~mg/mL;
    \item Gd\_2 --- an aqueous solution of gadopentetate dimeglumine (Magnevist\textsuperscript{\textregistered}, an MRI contrast agent) with a Gd concentration of 78.625~mg/mL;
    \item H$_2$O --- water;
    \item La\_1 --- composite contrast agent consisting of La$_2$O$_3$ nanoparticles stabilized on a matrix of few-layer graphite fragments (FGF) with a La concentration of 80~mg/mL;
    \item La\_2 --- aqueous solution of La(NO$_3$)$_3 \cdot 6$H$_2$O with a La concentration of 40~mg/mL;
    \item Nd\_1 --- composite contrast agent consisting of Nd$_2$O$_3$ nanoparticles stabilized on a matrix of few-layer graphite fragments (FGF) with a Nd concentration of 80~mg/mL;
    \item Nd\_2 --- aqueous solution of Nd(NO$_3$)$_3 \cdot 6$H$_2$O with an Nd concentration of 40~mg/mL;
    \item PMMA --- polymethyl methacrylate.
\end{itemize}

The Gd\_1, La\_1, and Nd\_1 samples were composite systems based on lanthanides and graphite/graphene fragments, conceptually similar to previously developed Ln--graphene contrast agents for PCCT~\cite{suslova2022nanomaterials}. In the present work, these variants were treated as separate classes, since their spectral responses may differ due to chemical form, concentration, matrix composition, or ROI extraction conditions.

The original dataset also contained an \texttt{Empty} class, corresponding to an empty or background region. This class was not included in the main material-classification task, because it does not represent a material in the same sense as the other classes. A separate experiment showed that including \texttt{Empty} mainly leads to confusion with \texttt{Air}; therefore, \texttt{Empty} was treated as an auxiliary background class.

\begin{figure}[htbp]
    \centering
    \includegraphics[width=0.72\textwidth]{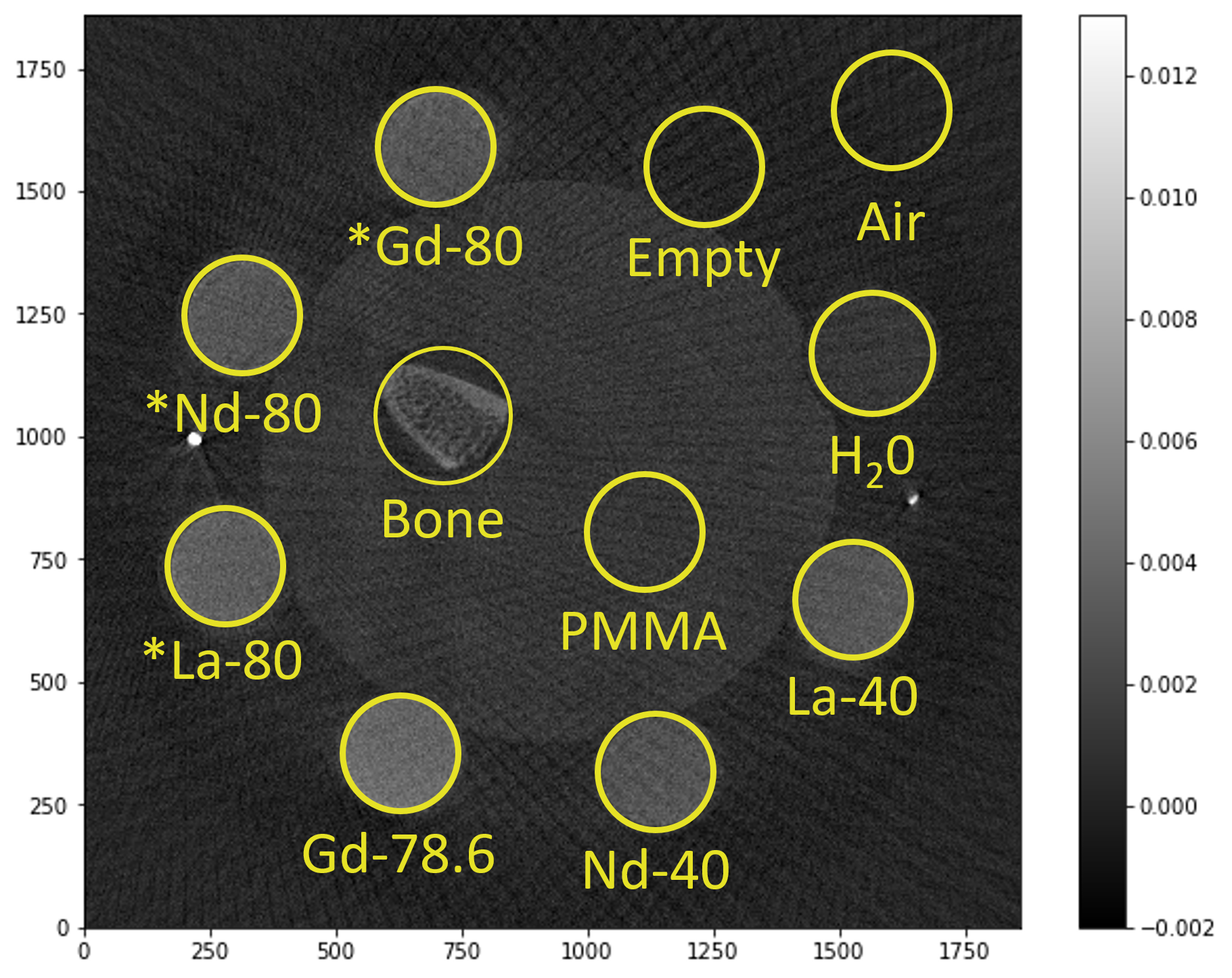}
    \caption{Example of a reconstructed phantom slice with marked ROIs for the investigated materials. Yellow circles indicate the regions used to extract the spectral features. The labels Gd-80, Gd-78.6, La-80, La-40, Nd-80, and Nd-40 correspond to the Gd\_1, Gd\_2, La\_1, La\_2, Nd\_1, and Nd\_2 classes, respectively. Asterisks preceding Gd-80, La-80, and Nd-80 indicate composite contrast-agent samples in which the corresponding lanthanide oxide nanoparticles were stabilized on a matrix of few-layer graphite fragments (FGF).}
    \label{fig:phantom_roi}
\end{figure}

An example of a reconstructed phantom slice with marked regions of interest is shown in figure~\ref{fig:phantom_roi}. Here, an ROI denotes an image region from which the spectral response of a material is extracted. This formulation differs from pixel-wise classification: the model is trained not on individual pixels, but on averaged spectra from image regions, which improves the statistical stability of the features.

\subsection{Data acquisition and reconstruction}
\label{subsec:data_acquisition}

The phantom was scanned using a laboratory photon-counting CT setup comprising a Source-Ray SB-120-350 X-ray source~\cite{sourceray_sb120350} and a silicon WidePIX hybrid-pixel photon-counting detector~\cite{jakubek2014widepix}. The X-ray tube was operated at 100~kVp and 150~$\mu$A. The source-to-detector distance was 350~mm.

A separate tomographic scan was acquired for each detector threshold setting. For each threshold, 360 angular projections were recorded. The exposure time was adjusted individually for each THL setting to compensate for the decrease in the photon count rate at higher thresholds. The exposure settings were selected so that the mean counting statistics per detector pixel remained approximately comparable over the investigated THL range. Consequently, longer exposure times were used at higher threshold values.

Before tomographic reconstruction, the raw projection images were processed using standard dark-field and flat-field corrections. The normalized transmission image was calculated as
\begin{equation}
    T =
    \frac{
        I_{\mathrm{raw}} - I_{\mathrm{dark}}
    }{
        I_{\mathrm{flat}} - I_{\mathrm{dark}}
    },
    \label{eq:flat_dark_correction}
\end{equation}
where $I_{\mathrm{raw}}$ is the measured projection, $I_{\mathrm{dark}}$ is the dark-field image acquired without X-ray exposure, and $I_{\mathrm{flat}}$ is the flat-field image acquired without the phantom. The corresponding line-integral projection data were calculated as
\begin{equation}
    p = -\ln T .
    \label{eq:log_projection}
\end{equation}

The corrected projection data acquired at each THL setting were reconstructed independently using the filtered back-projection (FBP) algorithm implemented in the ASTRA Toolbox~\cite{vanAarle2016astra}. The same acquisition geometry, reconstruction parameters, and output image grid were used for all threshold settings. This ensured spatial correspondence between the reconstructed datasets and allowed the same ROI locations to be analyzed throughout the threshold series.

The present analysis was performed using reconstructed threshold-dependent tomographic slices rather than direct transmission spectra. For each material, ROIs were extracted from corresponding spatial locations in the reconstructed images. The pixel values within each ROI were averaged, and the resulting mean reconstructed signal was used to form a spectral vector as a function of the detector threshold setting THL.

\subsection{Spectral features}
\label{subsec:spectral_features}

Experimentally, 12 detector threshold settings were used:
\begin{equation}
    \mathrm{THL}_{\mathrm{meas}} =
    \{45, 55, 65, 75, 85, 95, 105, 115, 125, 135, 145, 165\}.
    \label{eq:measured_thl}
\end{equation}

Each measured THL value corresponded to an independently acquired and reconstructed tomographic dataset. The THL values represent detector threshold settings and should not be interpreted directly as photon energies in keV without an additional detector-threshold calibration.

For subsequent model training, each measured ROI spectrum was interpolated onto a uniform THL grid extending from 45 to 165 with a step of one THL unit. The resulting interpolated spectral vector contained 121 values:
\begin{equation}
    \mathbf{I} =
    \left\{
        I_{45}, I_{46}, \ldots, I_{165}
    \right\},
    \label{eq:spectrum}
\end{equation}
where $I_k$ is the mean reconstructed signal within the ROI at the interpolated threshold value $k$.

Each dataset row additionally contained the number of pixels used to form the ROI spectrum and metadata describing the origin of the spectrum, including the ROI type, slice index, row index within the slice, source file, and subsampling information.

\begin{figure}[htbp]
    \centering
    \includegraphics[width=0.85\textwidth]{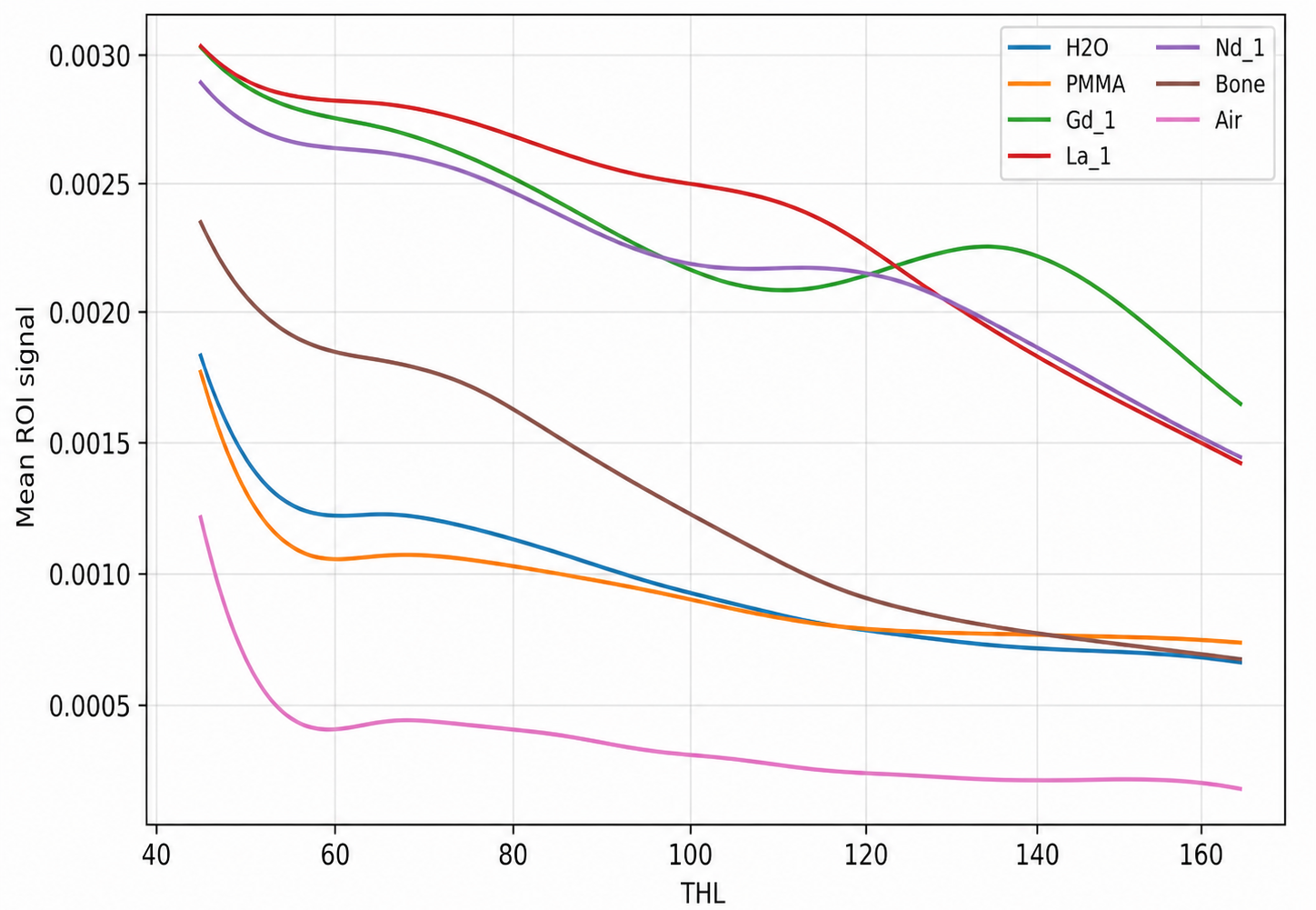}
    \caption{Examples of mean spectra for selected materials calculated from large ROIs. The horizontal axis shows the detector threshold setting THL, and the vertical axis shows the mean reconstructed signal within the ROI.}
    \label{fig:mean_spectra}
\end{figure}

Figure~\ref{fig:mean_spectra} shows examples of mean spectra for several material classes. Some materials exhibit clearly distinguishable spectral responses, whereas the H$_2$O/PMMA pair and several pairs of lanthanide-containing samples have more similar spectral shapes. These similarities are subsequently reflected in the confusion matrix.

\subsection{Large ROIs and random pixel subsamples}
\label{subsec:roi_types}

Two complementary ROI datasets were used. The \texttt{large} dataset was constructed from circular ROIs with radii of 90, 60, and 40 reconstructed-image pixels. For radii of 90 and 60 pixels, the dataset included one spectrum averaged over the complete circular ROI, one additional preprocessed ROI spectrum, and five spectra calculated from random subsets containing approximately one quarter of the pixels within the corresponding ROI. For the 40-pixel radius, the complete and preprocessed ROI spectra were used. This resulted in 16 spectra per material and tomographic slice in the \texttt{large} dataset.

The exact number of pixels contributing to each spectrum was stored in the dataset and used in the subsequent analysis.

The \texttt{small} dataset was generated by random pixel subsampling within the material ROIs. For each material and tomographic slice, 90 subsampled spectra were produced: 10 subsamples containing 3,000 pixels, 20 subsamples containing 1,000 pixels, 30 subsamples containing 500 pixels, and 30 subsamples containing 200 pixels. For each subsample, the corresponding pixel values were averaged at every THL setting to form a threshold-dependent spectral vector.

The random pixel subsamples were introduced to evaluate classifier robustness under reduced pixel statistics. Since multiple spectra derived from the same parent tomographic slice may be statistically correlated, all spectra sharing the same \texttt{slice\_id} were assigned to the same training, validation, or test subset, as described in section~\ref{sec:splitting}. The \texttt{all} dataset was formed by combining the \texttt{large} and \texttt{small} datasets.

\begin{table}[htbp]
\centering
\caption{ROI variants generated for each material and tomographic slice.}
\label{tab:roi_sets}
\begin{tabular}{llrr}
\toprule
Dataset & ROI or subsample type & Pixels or radius & Number per slice \\
\midrule
\texttt{large} & Complete circular ROI & radius 90 px & 1 \\
\texttt{large} & Preprocessed circular ROI & radius 90 px & 1 \\
\texttt{large} & Random quarter-ROI subsample & radius 90 px & 5 \\
\texttt{large} & Complete circular ROI & radius 60 px & 1 \\
\texttt{large} & Preprocessed circular ROI & radius 60 px & 1 \\
\texttt{large} & Random quarter-ROI subsample & radius 60 px & 5 \\
\texttt{large} & Complete circular ROI & radius 40 px & 1 \\
\texttt{large} & Preprocessed circular ROI & radius 40 px & 1 \\
\midrule
\texttt{small} & Random pixel subsample & 3000 pixels & 10 \\
\texttt{small} & Random pixel subsample & 1000 pixels & 20 \\
\texttt{small} & Random pixel subsample & 500 pixels & 30 \\
\texttt{small} & Random pixel subsample & 200 pixels & 30 \\
\bottomrule
\end{tabular}
\end{table}

\section{Dataset formation}
\label{sec:dataset}

The input files were converted into a unified tabular structure, where each row corresponded to one ROI spectrum. For each row, the following metadata were stored:
\begin{enumerate}
    \item material class --- the label to be predicted by the model, for example \texttt{Air}, \texttt{Bone}, Gd\_1, H$_2$O, or PMMA;
    
    \item ROI type --- an indication of whether the spectrum was obtained from a large ROI (\texttt{large}) or from a small ROI/random pixel subsample (\texttt{small}); the combined dataset \texttt{all} was formed as the union of rows of both types and does not represent a separate physical ROI type;
    
    \item source file name --- the name of the file from which the spectrum was obtained. This makes it possible to trace the origin of each dataset row and, if necessary, return to the original data;
    
    \item slice index --- the number of the tomographic slice from which the ROI was extracted. This parameter was used for the group split: all ROIs obtained from the same slice were assigned to only one subset;
    
    \item row index within the slice --- a technical number of the ROI or subsample within a given slice. This index is needed to distinguish multiple spectra obtained from the same tomographic slice;
    
    \item number of pixels in the ROI --- the number of pixels over which the signal was averaged when forming the spectrum. A larger number of pixels corresponds to higher counting statistics and a more stable spectral vector;
    
    \item ROI or subsampling parameters --- geometric or technical parameters of the region of interest, such as the ROI center coordinates, radius, region size, or random pixel subsampling parameters;
    
    \item spectral features for THL 45--165 --- numerical values of the ROI signal at different THL after interpolation onto a uniform grid. These features were used as the input vector for the neural-network model.
\end{enumerate}

The main formulation used a 10-class task without the \texttt{Empty} class. The final dataset contained 234144 rows. The split into training, validation, and test subsets used in the final experiment is shown in table~\ref{tab:split}.

\begin{table}[htbp]
\centering
\caption{Dataset split used in the final experiment.}
\label{tab:split}
\begin{tabular}{lrr}
\toprule
Subset & Number of ROI spectra & Number of slice groups \\
\midrule
Training   & 148508 & 137 \\
Validation & 37940  & 35  \\
Test       & 47696  & 44  \\
\midrule
Total      & 234144 & 216 \\
\bottomrule
\end{tabular}
\end{table}

\section{Data splitting and leakage prevention}
\label{sec:splitting}

A key feature of the dataset is the presence of correlated observations. Multiple spectra were generated from each reconstructed tomographic slice, including large-ROI spectra and multiple small-ROI subsamples. A conventional random row-wise split could therefore assign strongly correlated spectra derived from the same slice to different subsets, resulting in an overly optimistic estimate of classification performance.

To reduce this risk, the main evaluation used a group split based on the integer \texttt{slice\_id}. During dataset construction, \texttt{slice\_id} ranged from 0 to 215 and was assigned consistently across all source files. Consequently, all ROI spectra sharing the same \texttt{slice\_id}, including spectra from different material classes, source files, ROI sizes, and subsampling variants, were assigned to the same subset.

The dataset contained 216 unique slice groups. A two-stage \texttt{GroupShuffleSplit} procedure with a random seed of 42 assigned 137 groups to the training subset, 35 groups to the validation subset, and 44 groups to the test subset. The corresponding numbers of ROI spectra were 148,508, 37,940, and 47,696, respectively. The sets of \texttt{slice\_id} values assigned to the training, validation, and test subsets were mutually disjoint.

Feature standardization was performed only after the data had been split. The parameters of the \texttt{StandardScaler} were estimated exclusively from the training subset and were subsequently applied unchanged to the validation and test subsets.

\section{Spectral preprocessing}
\label{sec:preprocessing}

Several feature representations were evaluated.

\subsection{Raw spectra}

The \texttt{raw} representation used the original interpolated spectrum:
\begin{equation}
    \mathbf{X} = \mathbf{I} .
    \label{eq:raw}
\end{equation}

This representation preserves both the spectral shape and the absolute signal level.

\subsection{Logarithmic transformation}

The \texttt{log1p} representation used the transformation
\begin{equation}
    \mathbf{X} = \log(1 + \mathbf{I}) .
    \label{eq:log1p}
\end{equation}

This transformation reduces the effect of the large dynamic range of the measured intensities.

\subsection{Area normalization}

The \texttt{area} representation normalized the spectrum by its integral intensity:
\begin{equation}
    \mathbf{X} = \frac{\mathbf{I}}{\sum_k I_k} .
    \label{eq:area}
\end{equation}

This representation emphasizes the spectral shape while suppressing the absolute amplitude.

\subsection{Logarithmic spectrum with derivative}

The \texttt{log1p\_deriv} representation was defined as the concatenation of the logarithmically transformed spectrum and its first derivative with respect to THL:
\begin{equation}
    \mathbf{X} = \left[\log(1 + \mathbf{I}),\; \frac{d}{d\mathrm{THL}} \log(1 + \mathbf{I})\right] .
    \label{eq:log1p_deriv}
\end{equation}

The original spectrum contained 121 points; after combining the transformed spectrum with its derivative, the input feature vector contained 242 features. This representation preserves information about the signal level while additionally incorporating information about local changes in the spectral curve.

\begin{figure}[htbp]
    \centering
    \includegraphics[width=0.85\textwidth]{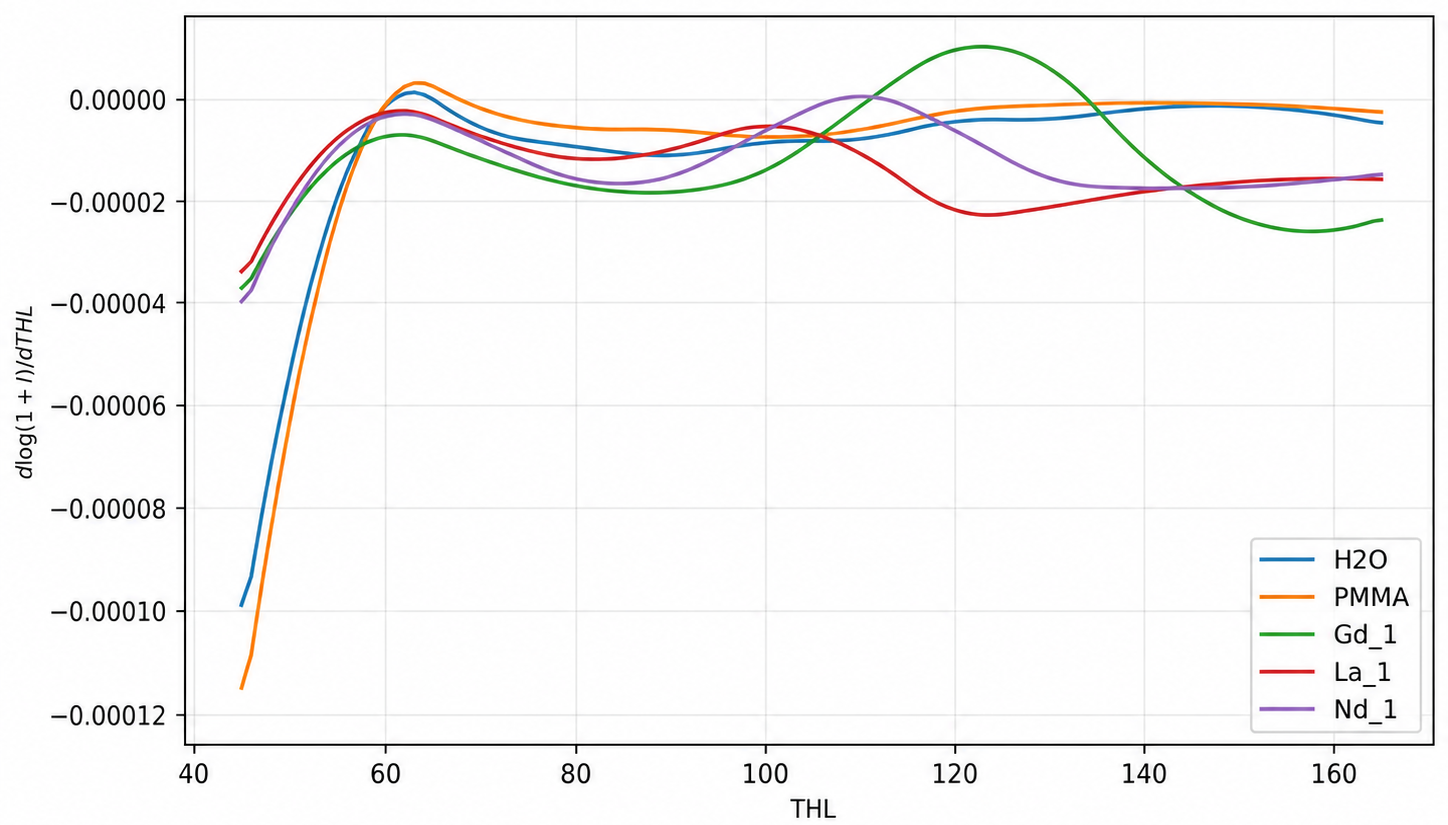}
    \caption{Example derivative features $d\log(1+I)/d\mathrm{THL}$ for selected materials. These features were used together with the logarithmically transformed spectrum in the \texttt{log1p\_deriv} representation.}
    \label{fig:derivative_features}
\end{figure}

An example of derivative features for selected classes is shown in figure~\ref{fig:derivative_features}. The derivative enhances local changes in the spectral shape and therefore complements the absolute signal values.

\section{Neural-network models}
\label{sec:models}

Two neural-network architectures were compared: a multilayer perceptron and a one-dimensional convolutional neural network.

\begin{figure}[htbp]
    \centering
    \includegraphics[width=0.9\textwidth]{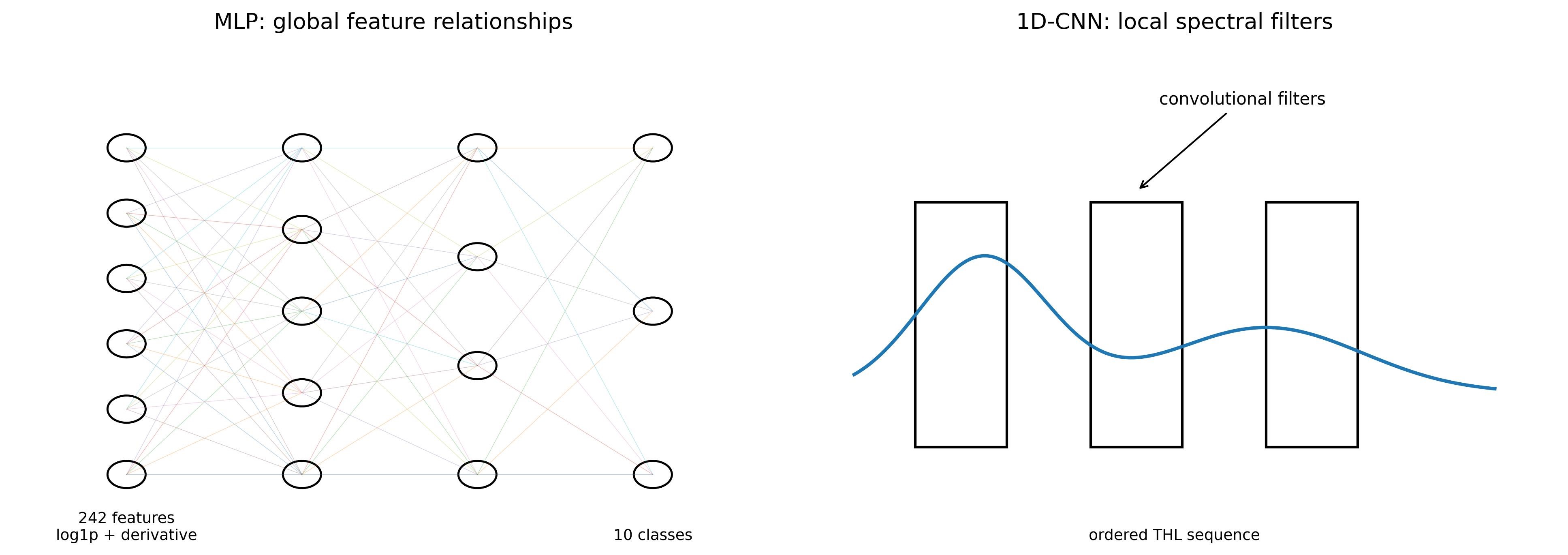}
    \caption{Conceptual difference between the MLP and 1D-CNN architectures.}
    \label{fig:mlp_cnn_scheme}
\end{figure}

The conceptual difference between these architectures is shown in figure~\ref{fig:mlp_cnn_scheme}. The MLP treats the spectral vector as a set of interrelated features and can use global relationships between all THL values. In contrast, the 1D-CNN applies local filters along the THL axis and is therefore oriented toward detecting local patterns in the spectrum.

\begin{table}[htbp]
\centering
\small
\caption{Architecture of the neural-network models used in this work.}
\label{tab:model_architecture}
\begin{tabularx}{\textwidth}{p{2.0cm} >{\raggedright\arraybackslash}X}
\toprule
Model & Architecture \\
\midrule
MLP &
Linear(242, 256) -- BatchNorm -- ReLU -- Dropout(0.25); \newline
Linear(256, 128) -- BatchNorm -- ReLU -- Dropout(0.20); \newline
Linear(128, 64) -- ReLU; \newline
Linear(64, 10). \\
\midrule
1D-CNN &
Conv1d(1, 32, kernel=5) -- BatchNorm -- ReLU -- MaxPool; \newline
Conv1d(32, 64, kernel=5) -- BatchNorm -- ReLU -- MaxPool; \newline
Conv1d(64, 128, kernel=3) -- BatchNorm -- ReLU -- AdaptiveAvgPool; \newline
Linear(128, 64) -- ReLU -- Linear(64, 10). \\
\bottomrule
\end{tabularx}
\end{table}

\begin{table}[htbp]
\centering
\small
\caption{Main training parameters of the neural-network models.}
\label{tab:training_parameters}
\begin{tabularx}{\textwidth}{p{5.0cm} >{\raggedright\arraybackslash}X}
\toprule
Parameter & Value \\
\midrule
Optimizer & AdamW \\
Learning rate & 0.001 \\
Weight decay & 0.0001 \\
Loss function & CrossEntropyLoss \\
Batch size & 512 \\
Maximum number of epochs & 30 \\
Scheduler & ReduceLROnPlateau, factor 0.5, patience 5 \\
Early stopping patience & 12 \\
Feature scaling & StandardScaler \\
Random seed & 42 \\
\bottomrule
\end{tabularx}
\end{table}

\subsection{Multilayer perceptron}
\label{subsec:mlp}

The neural-network models were implemented using PyTorch~\cite{paszke2019pytorch}; data splitting and metric calculation were performed using scikit-learn~\cite{pedregosa2011scikit}. The AdamW optimizer with decoupled weight decay was used for training~\cite{loshchilov2019adamw}.

The multilayer perceptron treats the spectral vector as a set of numerical features. It does not impose an explicit assumption of locality along the THL axis, but it can use global relationships between all spectral points.

\subsection{One-dimensional convolutional neural network}
\label{subsec:cnn}

The 1D-CNN treats the spectrum as an ordered sequence in which neighboring THL values are related to each other. Convolutional filters are applied along the THL axis to extract local patterns in the spectral curve. This architecture therefore introduces an explicit locality assumption, in contrast to the fully connected MLP architecture.

\section{Results}
\label{sec:results}

\subsection{Comparison of MLP and 1D-CNN}
\label{subsec:model_comparison}

Three metrics were used to evaluate classification performance: accuracy, balanced accuracy, and macro-averaged F1-score. Accuracy shows the fraction of all correctly classified ROI spectra in the test set:
\begin{equation}
    \mathrm{Accuracy} =
    \frac{N_{\mathrm{correct}}}{N_{\mathrm{all}}},
\end{equation}
where $N_{\mathrm{correct}}$ is the number of correctly classified objects, and $N_{\mathrm{all}}$ is the total number of objects in the test set.

However, in multi-class classification, accuracy alone may be insufficient. If one class is represented much more frequently than the others, high accuracy may be achieved mainly by correctly recognizing the most numerous classes. Therefore, balanced accuracy was additionally used. This metric first calculates the recall for each class and then averages these values over all classes:
\begin{equation}
    \mathrm{Balanced~accuracy} =
    \frac{1}{C}
    \sum_{c=1}^{C}
    \frac{TP_c}{TP_c + FN_c},
\end{equation}
where $C$ is the number of classes, $TP_c$ is the number of objects of class $c$ correctly assigned to this class, and $FN_c$ is the number of objects of class $c$ incorrectly assigned to other classes. Thus, balanced accuracy gives equal weight to each class regardless of the number of objects in it.

The third metric was the macro-averaged F1-score. For each class, precision, recall, and F1-score were calculated separately:
\begin{equation}
    \mathrm{Precision}_c =
    \frac{TP_c}{TP_c + FP_c},
\end{equation}
\begin{equation}
    \mathrm{Recall}_c =
    \frac{TP_c}{TP_c + FN_c},
\end{equation}
\begin{equation}
    F1_c =
    2
    \frac{
    \mathrm{Precision}_c \cdot \mathrm{Recall}_c
    }{
    \mathrm{Precision}_c + \mathrm{Recall}_c
    }.
\end{equation}
The $F1_c$ values were then averaged over all classes:
\begin{equation}
    \mathrm{Macro~F1} =
    \frac{1}{C}
    \sum_{c=1}^{C} F1_c.
\end{equation}

In the present work, the accuracy, balanced accuracy, and macro-averaged F1-score of the final model were almost identical and amounted to approximately 0.9745. This is due to two factors. First, the test set in the main 10-class formulation was sufficiently balanced in terms of the number of ROI spectra for different materials. Second, the recognition quality was not concentrated only on one or several classes: the model showed high F1-scores for most materials, while the remaining errors were distributed among a limited number of spectrally close pairs, such as H$_2$O/PMMA, Gd\_1/Gd\_2, and several La/Nd-containing samples. Therefore, the overall fraction of correct predictions, the average class recall, and the average class F1-score gave similar numerical values.

The closeness of these three metrics is important for interpreting the result. It shows that the high accuracy of the final model is not a consequence of one well-recognized dominant class, but reflects stable classification performance across most material classes.

The initial comparison of the two architectures was performed for large ROIs, small ROIs, and the combined dataset. The MLP consistently outperformed the 1D-CNN, as shown in table~\ref{tab:model_comparison}. This comparison was performed using the \texttt{raw} spectral representation in order to compare the two architectures under the same baseline feature representation.

\begin{table}[htbp]
\centering
\caption{Comparison of MLP and 1D-CNN for different ROI datasets using the \texttt{raw} spectral representation.}
\label{tab:model_comparison}
\begin{tabular}{llccc}
\hline
Model & ROI dataset & Accuracy & Balanced accuracy & Macro F1 \\
\hline
1D-CNN & all   & 0.9465 & 0.9460 & 0.9461 \\
1D-CNN & large & 0.9941 & 0.9935 & 0.9936 \\
1D-CNN & small & 0.9519 & 0.9519 & 0.9520 \\
MLP    & all   & 0.9651 & 0.9649 & 0.9649 \\
MLP    & large & 0.9995 & 0.9996 & 0.9996 \\
MLP    & small & 0.9772 & 0.9772 & 0.9772 \\
\hline
\end{tabular}
\end{table}

For large ROIs, both models achieved high performance; however, the MLP approached nearly perfect classification. The advantage of the MLP was more pronounced for small ROIs, indicating better robustness to spectra with limited statistics.

\begin{figure}[htbp]
    \centering
    \includegraphics[width=0.82\textwidth]{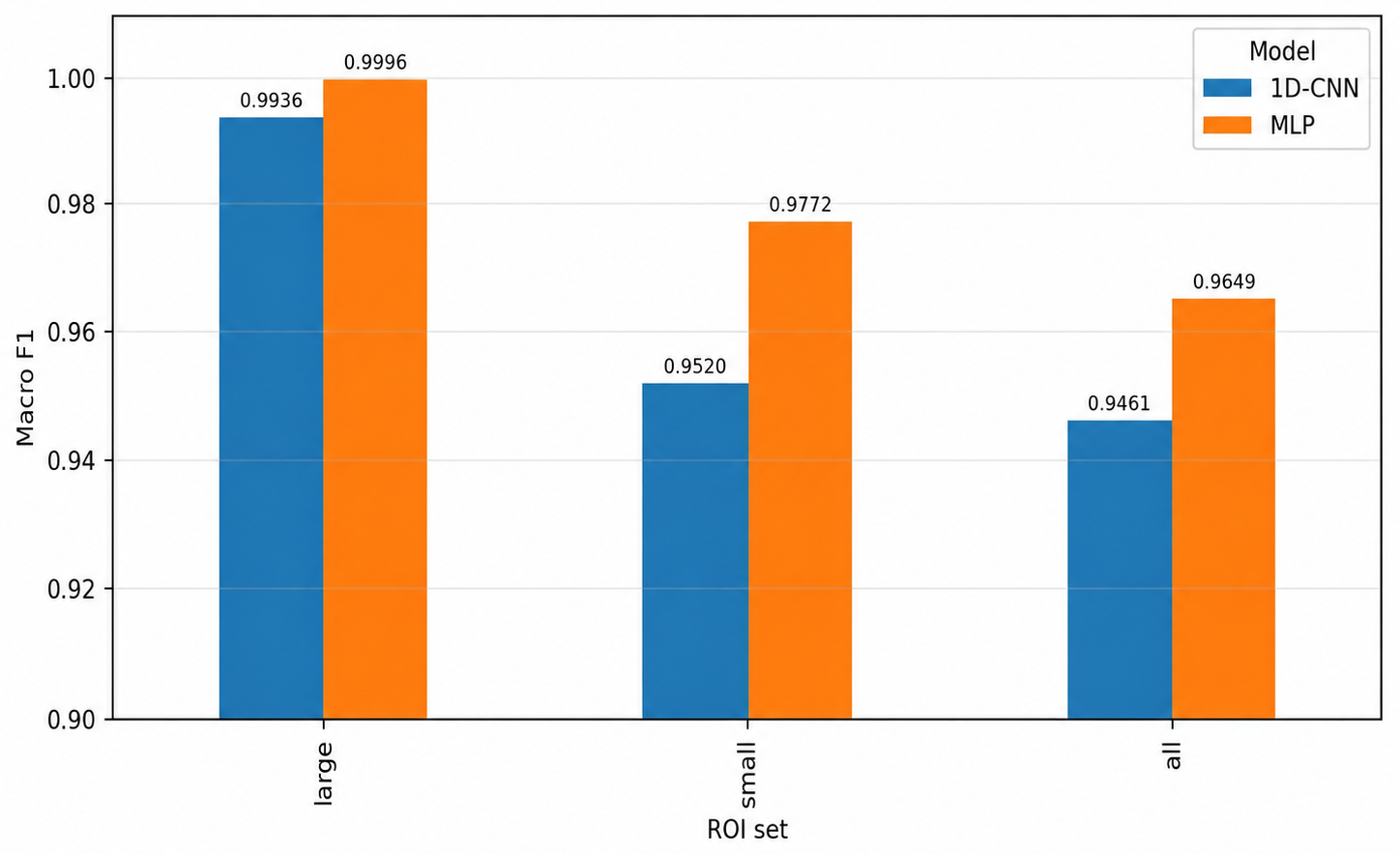}
    \caption{Comparison of MLP and 1D-CNN in terms of macro F1 for large ROIs, small ROIs, and the combined dataset.}
    \label{fig:model_f1}
\end{figure}

A graphical comparison of macro F1 for the two architectures is shown in figure~\ref{fig:model_f1}. For large ROIs, the difference between the models is small because the spectra have high statistics. For small ROIs and for the combined dataset, the MLP provides higher performance. This indicates that in the present task not only local spectral patterns but also global amplitude relationships between features play an important role.

\subsection{Comparison of preprocessing strategies}
\label{subsec:preproc_comparison}

Several preprocessing strategies were compared for the combined ROI dataset using the MLP architecture. The results are shown in table~\ref{tab:preprocessing}.

\begin{table}[htbp]
\centering
\caption{Effect of spectral preprocessing on MLP performance for the combined ROI dataset.}
\label{tab:preprocessing}
\begin{tabular}{lcccc}
\hline
Preprocessing & Number of features & Accuracy & Balanced accuracy & Macro F1 \\
\hline
\texttt{area}          & 121 & 0.8912 & 0.8912 & 0.8908 \\
\texttt{raw}           & 121 & 0.9590 & 0.9588 & 0.9587 \\
\texttt{log1p}         & 121 & 0.9613 & 0.9612 & 0.9610 \\
\texttt{log1p\_deriv} & 242 & 0.9745 & 0.9745 & 0.9745 \\
\hline
\end{tabular}
\end{table}

Area normalization substantially reduced the classification performance. This shows that the absolute intensity contains useful information for distinguishing materials. The logarithmic transformation provided a moderate improvement compared with the raw representation, while adding the derivative led to the most pronounced performance gain.

\begin{figure}[htbp]
    \centering
    \includegraphics[width=0.75\textwidth]{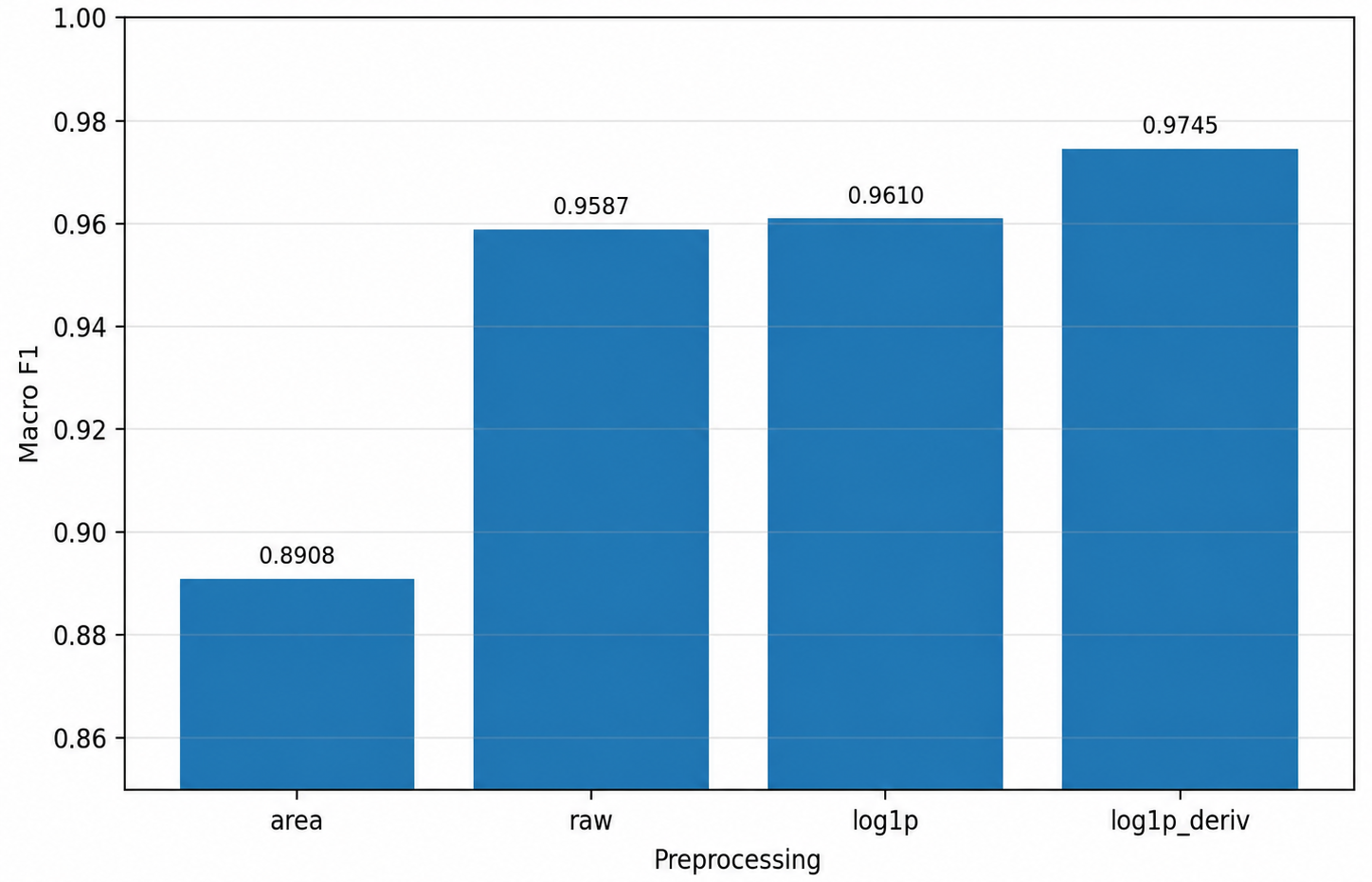}
    \caption{Effect of spectral preprocessing on macro F1 for the MLP on the combined ROI dataset.}
    \label{fig:preprocessing_f1}
\end{figure}

Figure~\ref{fig:preprocessing_f1} shows that the \texttt{log1p\_deriv} representation is the most effective for the combined ROI dataset. At the same time, area normalization of the spectrum leads to a sharp decrease in performance, indicating that the absolute signal level is not a nuisance scaling factor, but contains information useful for material identification.

\subsection{Final model}
\label{subsec:final_model}

The final model used the MLP architecture, group splitting, the 10-class formulation without \texttt{Empty}, the combined ROI dataset, and the \texttt{log1p\_deriv} representation. The configuration of the final model is shown in table~\ref{tab:final_configuration}.

\begin{table}[htbp]
\centering
\caption{Configuration of the final model.}
\label{tab:final_configuration}
\begin{tabular}{ll}
\hline
Parameter & Value \\
\hline
Architecture & MLP \\
Splitting & group split by slice index \\
Classes & 10, without \texttt{Empty} \\
ROI dataset & \texttt{all}, combination of \texttt{large} and \texttt{small} ROIs \\
Preprocessing & \texttt{log1p\_deriv} \\
Number of rows & 234144 \\
Number of features & 242 \\
Random seed & 42 \\
Number of epochs & 30 \\
Best epoch & 30 \\
\hline
\end{tabular}
\end{table}

The final metrics on the test set are shown in table~\ref{tab:final_metrics}.

\begin{table}[htbp]
\centering
\caption{Performance of the final model on the test set.}
\label{tab:final_metrics}
\begin{tabular}{lc}
\hline
Metric & Value \\
\hline
Accuracy & 0.9745 \\
Balanced accuracy & 0.9745 \\
Macro F1 & 0.9745 \\
\hline
\end{tabular}
\end{table}

The validation macro F1-score increased from 0.9458 at epoch 1 to 0.9752 at epoch 30. Since the highest validation score was obtained at the final training epoch, convergence beyond 30 epochs was not assessed.

\subsection{Class-wise performance}
\label{subsec:classwise}

The precision, recall, and F1-score values of the final model for individual classes are presented in table~\ref{tab:classwise}.

\begin{table}[htbp]
\centering
\caption{Class-wise performance of the final MLP model.}
\label{tab:classwise}
\begin{tabular}{lcccc}
\hline
Class & Precision & Recall & F1-score & Support \\
\hline
\texttt{Air}  & 1.00 & 1.00 & 1.00 & 5192 \\
\texttt{Bone} & 1.00 & 1.00 & 1.00 & 4664 \\
Gd\_1 & 0.97 & 0.98 & 0.98 & 4664 \\
Gd\_2 & 0.98 & 0.96 & 0.97 & 4664 \\
H$_2$O  & 0.94 & 0.94 & 0.94 & 4664 \\
La\_1 & 0.97 & 0.98 & 0.97 & 4664 \\
La\_2 & 1.00 & 1.00 & 1.00 & 4664 \\
Nd\_1 & 0.97 & 0.95 & 0.96 & 4664 \\
Nd\_2 & 0.97 & 0.99 & 0.98 & 4664 \\
PMMA & 0.95 & 0.95 & 0.95 & 5192 \\
\hline
\end{tabular}
\end{table}

The most robustly recognized classes were \texttt{Air}, \texttt{Bone}, and La\_2, for which the F1-score reached 1.00. The main residual errors were associated with H$_2$O and PMMA, as well as with some specific pairs of Ln-based samples.

\begin{figure}[htbp]
    \centering
    \includegraphics[width=0.82\textwidth]{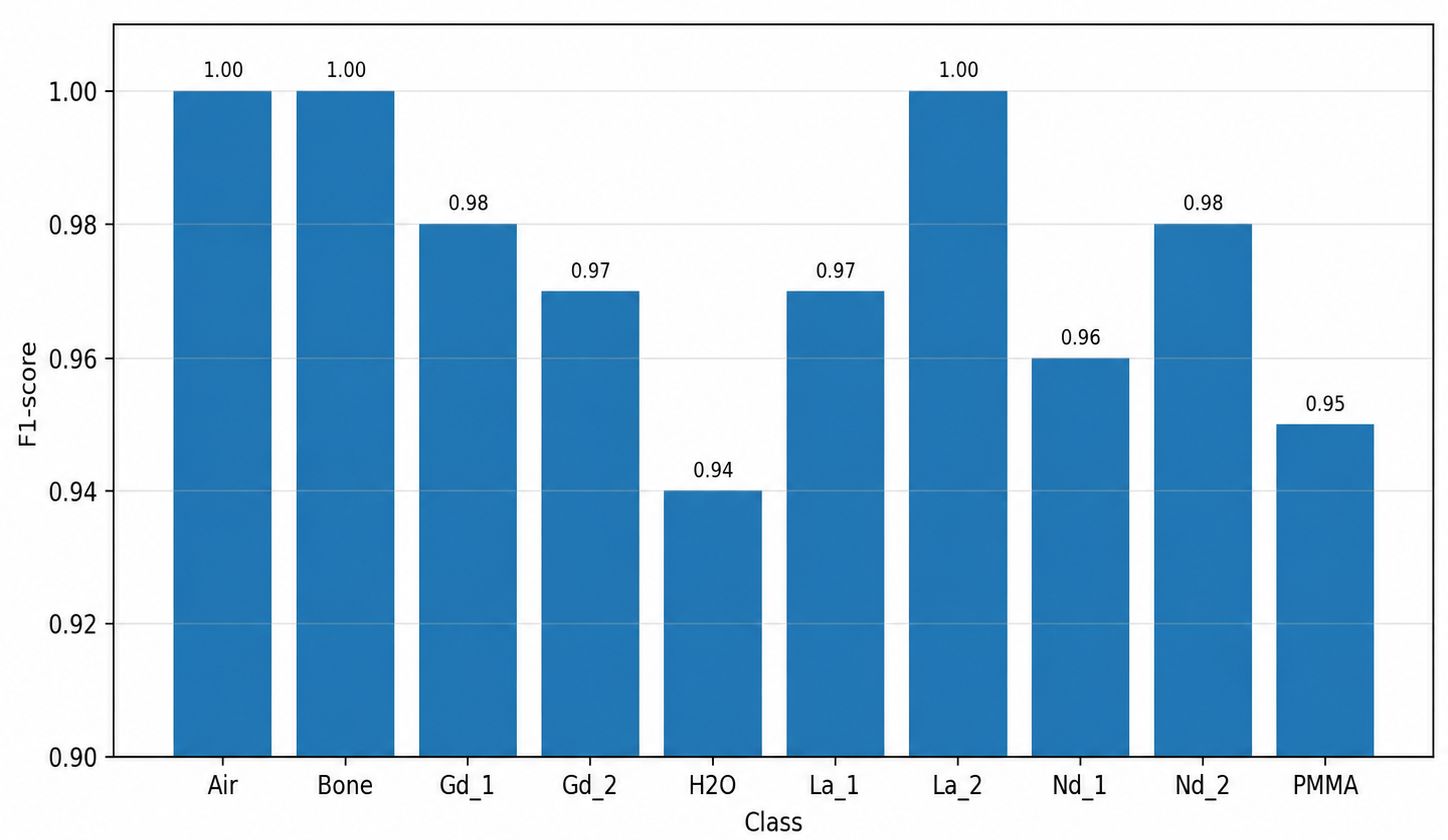}
    \caption{Class-wise F1-score for the final MLP model. The most challenging classes remain H$_2$O, PMMA, and Nd\_1.}
    \label{fig:classwise_f1}
\end{figure}

\begin{figure}[htbp]
    \centering
    \includegraphics[width=0.78\textwidth]{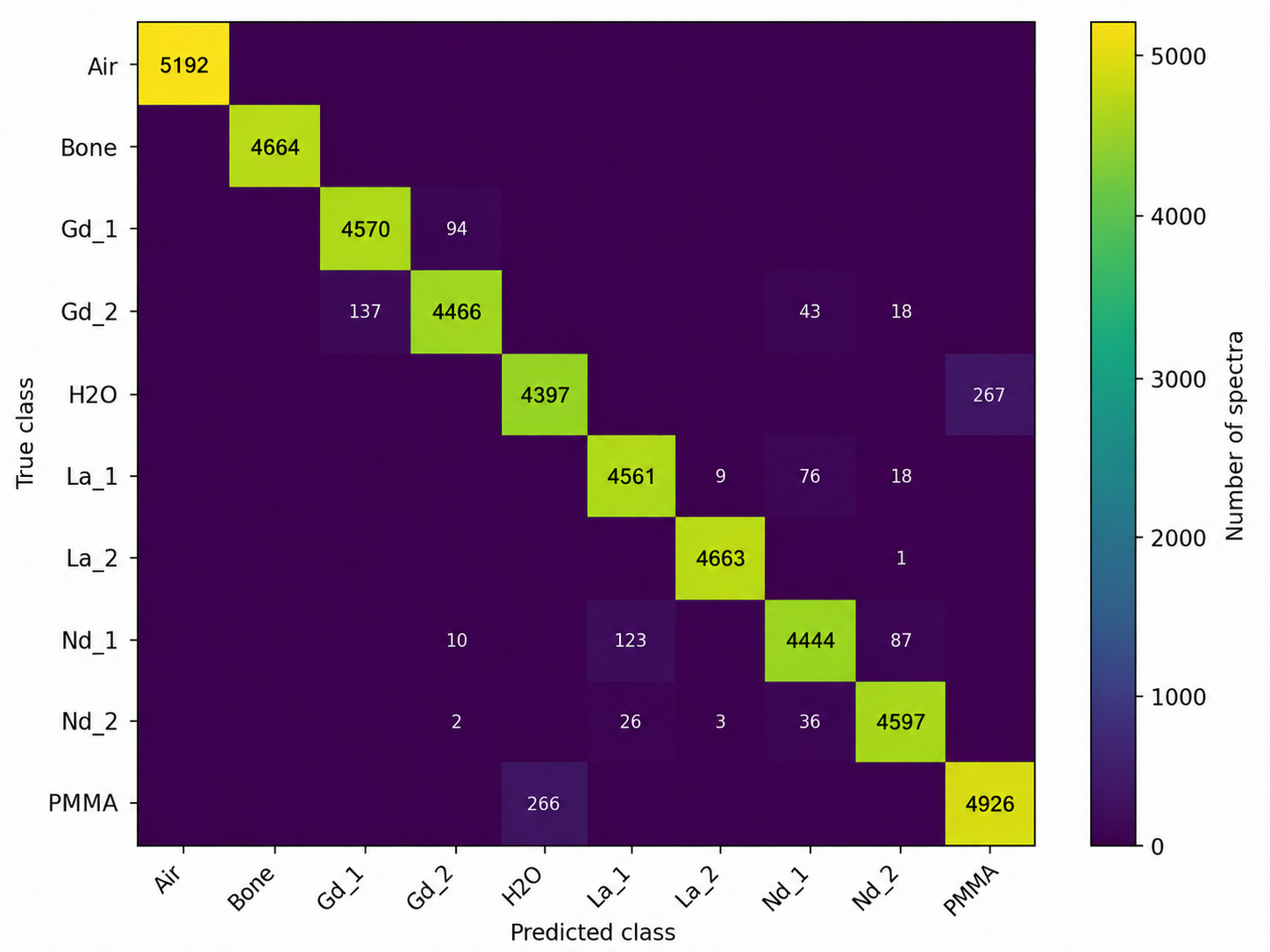}
    \caption{Confusion matrix of the final MLP model for the 10-class formulation without \texttt{Empty}. The main confusion is associated with H$_2$O/PMMA, Gd\_1/Gd\_2, and several La/Nd-containing samples.}
    \label{fig:confusion_final}
\end{figure}

Figure~\ref{fig:classwise_f1} shows the distribution of F1-scores by class, and figure~\ref{fig:confusion_final} shows the confusion matrix of the final MLP model. The dominant errors are quantified in table~\ref{tab:dominant_errors}. The most pronounced confusion was observed between H$_2$O and PMMA: 267 H$_2$O spectra were classified as PMMA, while 266 PMMA spectra were classified as H$_2$O. This nearly symmetric confusion indicates the similarity of their spectral responses over the considered THL range, consistent with the absence of pronounced absorption edges and with their relatively similar effective attenuation behavior.

\begin{table}[htbp]
\centering
\caption{Dominant errors of the final MLP model according to the confusion matrix.}
\label{tab:dominant_errors}
\begin{tabular}{llr}
\hline
True class & Predicted class & Number of errors \\
\hline
H$_2$O & PMMA & 267 \\
PMMA & H$_2$O & 266 \\
Gd\_2 & Gd\_1 & 137 \\
Nd\_1 & La\_1 & 123 \\
Gd\_1 & Gd\_2 & 94 \\
Nd\_1 & Nd\_2 & 87 \\
La\_1 & Nd\_1 & 76 \\
Gd\_2 & Nd\_1 & 43 \\
\hline
\end{tabular}
\end{table}

\begin{figure}[htbp]
    \centering
    \includegraphics[width=0.82\textwidth]{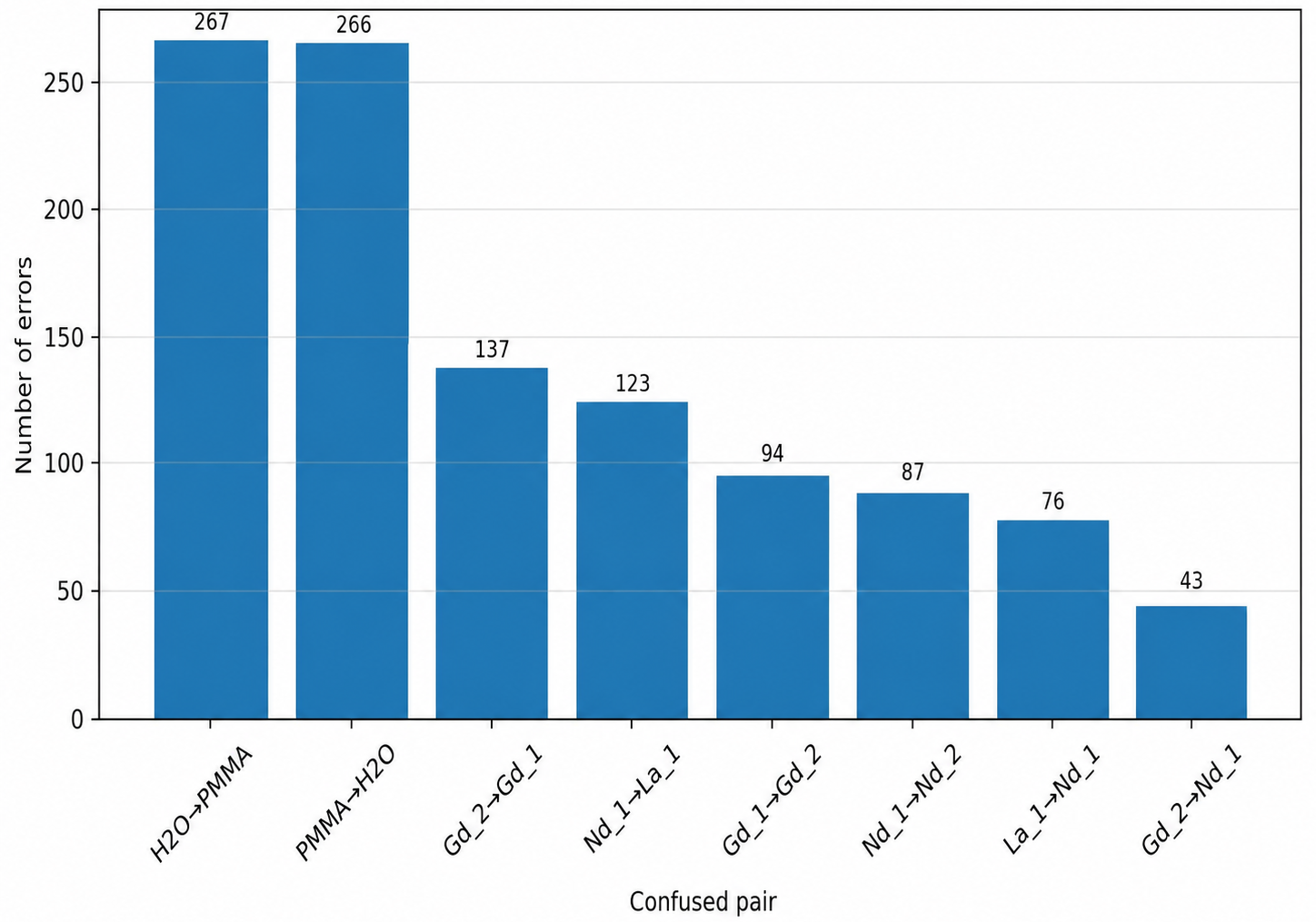}
    \caption{Most frequent misclassification pairs of the final MLP model.}
    \label{fig:error_pairs}
\end{figure}

\subsection{Separate analysis of the Gd\_1 and Gd\_2 classes}
\label{subsec:gd_pair_analysis}

The Gd\_1/Gd\_2 pair deserves separate discussion. These classes may appear difficult to distinguish, since both samples contain gadolinium at comparable concentrations (80~mg/mL for Gd\_1 and 78.625~mg/mL for Gd\_2). However, the experimental analysis showed that their ROI spectra differ systematically.

\begin{figure}[htbp]
    \centering
    \includegraphics[width=0.82\textwidth]{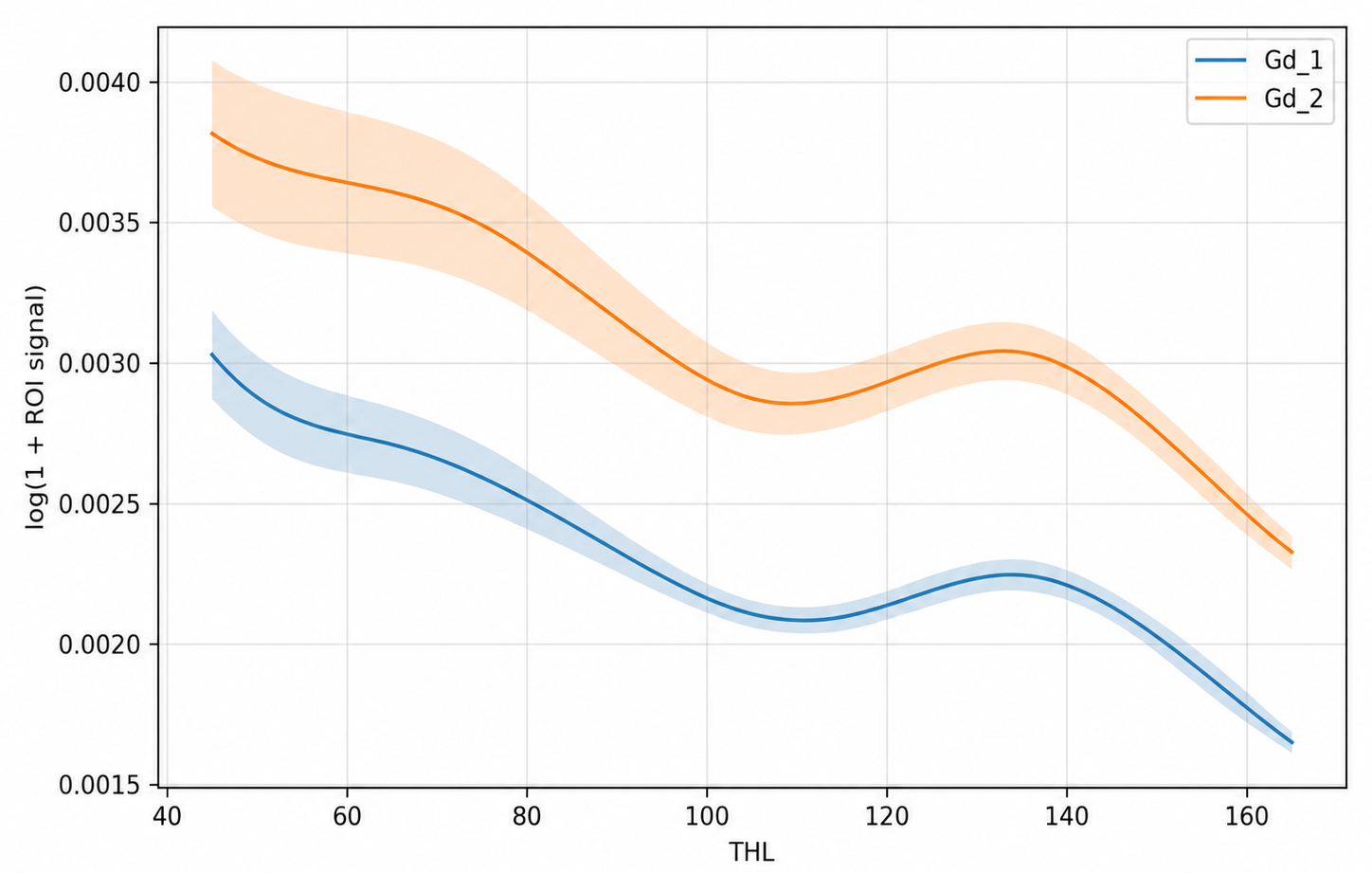}
    \caption{
    Comparison of the mean log1p spectra for the Gd\_1 and Gd\_2 classes calculated from large ROIs. The shaded regions indicate $\pm 1$ standard deviation across the analyzed large-ROI spectra.
    }
    \label{fig:gd_log1p_mean}
\end{figure}

\begin{figure}[htbp]
    \centering
    \includegraphics[width=0.82\textwidth]{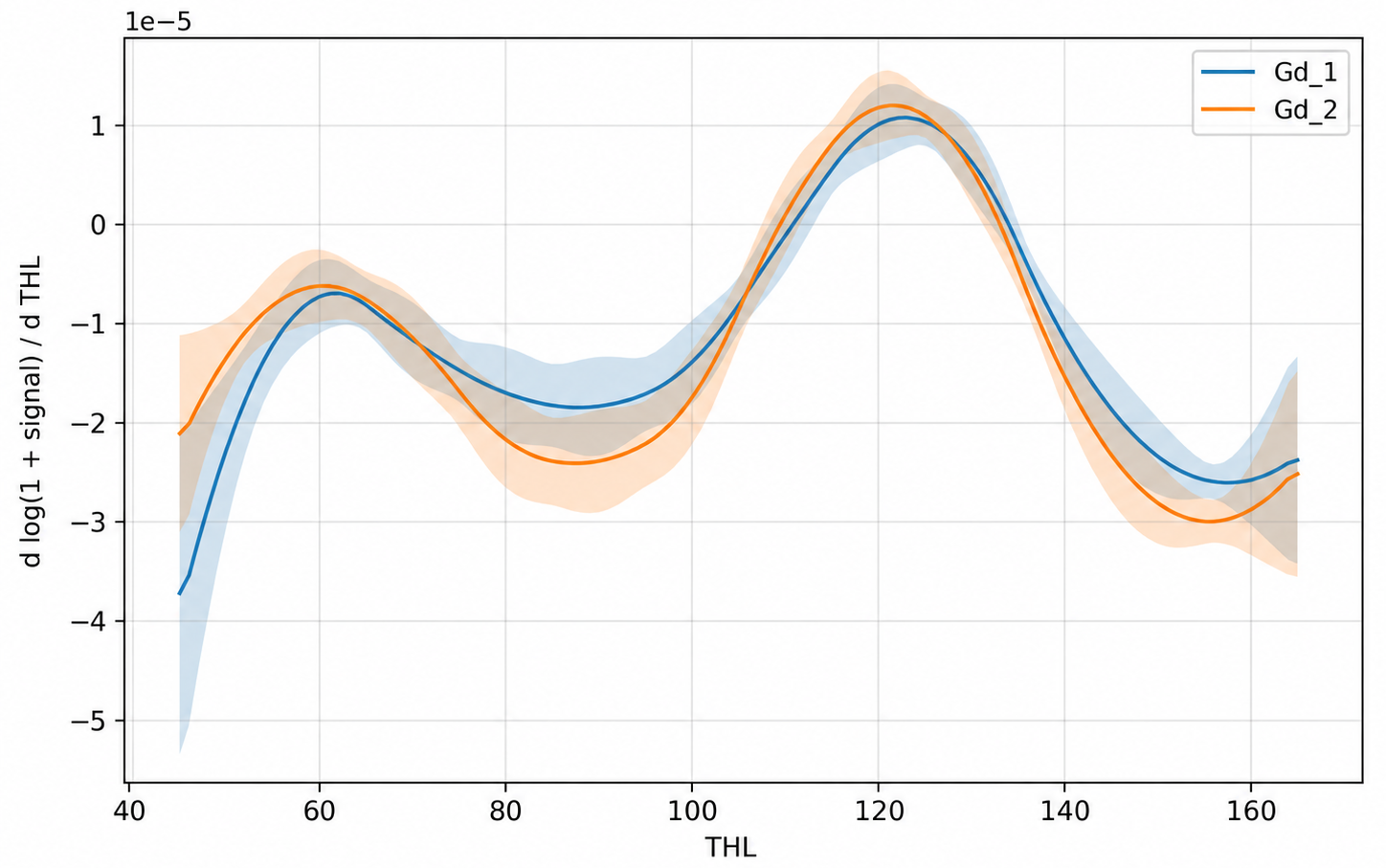}
    \caption{
    Comparison of the mean log1p-derivative spectra for the Gd\_1 and Gd\_2 classes. The shaded regions indicate $\pm 1$ standard deviation across the analyzed large-ROI spectra.
    }
    \label{fig:gd_log1p_derivative}
\end{figure}

For large ROIs, both classes were represented symmetrically: 3456 spectra, 216 tomographic slices, and one source file for each class. Comparison of the mean log1p spectra showed that the Gd\_2 curve lies above the Gd\_1 curve over almost the entire THL range, while the Gd\_1--Gd\_2 difference retains the same sign throughout the interval (Figure~\ref{fig:gd_log1p_mean}). This indicates a stable difference in the full spectral response of the two samples. Comparison of the log1p-derivative spectra additionally shows differences in the local shape and slope of the spectral curves (Figure~\ref{fig:gd_log1p_derivative}).

In the test set, the model correctly classified 4570 Gd\_1 objects and 4466 Gd\_2 objects. The main confusion occurred within the pair: 94 Gd\_1 objects were assigned to Gd\_2, and 137 Gd\_2 objects were assigned to Gd\_1. In addition, several Gd\_2 errors were observed as Gd\_2$\rightarrow$Nd\_1 and Gd\_2$\rightarrow$Nd\_2, with 43 and 18 cases, respectively.

\begin{table}[htbp]
\centering
\caption{Classification outcomes for the true Gd\_1 and Gd\_2 classes in the test set.}
\label{tab:gd_errors}
\begin{tabular}{llr}
\toprule
True class & Predicted class & Number of objects \\
\midrule
Gd\_1 & Gd\_1 & 4570 \\
Gd\_1 & Gd\_2 & 94 \\
Gd\_2 & Gd\_2 & 4466 \\
Gd\_2 & Gd\_1 & 137 \\
Gd\_2 & Nd\_1 & 43 \\
Gd\_2 & Nd\_2 & 18 \\
\bottomrule
\end{tabular}
\end{table}

Thus, the model distinguishes these two classes with high but not perfect accuracy. This result should not be interpreted as the ability of the model to determine the Gd concentration with an accuracy of about two percent. A more appropriate interpretation is that the neural network distinguishes two specific experimental samples represented in the training dataset. Gd\_1 is a sample based on nanoparticles stabilized on few-layer graphite fragments, whereas Gd\_2 corresponds to an aqueous solution of a Gd-based molecular contrast agent. Therefore, the differences between the classes may be related not only to gadolinium concentration, but also to the sample matrix, solvent, density, physicochemical form of the compound, and the overall spectral signal level.

This distinction separates the present formulation from tasks focused only on detecting the presence of Gd as an element. In this work, the classifier is trained to distinguish specific sample classes using the full ROI spectrum. Therefore, it can exploit weak but reproducible differences in the shape of the spectral curve that are not necessarily reducible to the K-edge feature of a single element. At the same time, transferring this result to new Gd concentrations, other solvents, or other matrices requires separate experimental validation.

\subsection{Comparison with the gradient-boosting baseline}
\label{subsec:boosting_baseline}

Preliminary results for this task were obtained using gradient-boosted decision trees, including XGBoost, LightGBM, and CatBoost. In that formulation, the 12 original THL were used, ROIs were extracted followed by averaging of pixel values, and a standard row-wise \texttt{train\_test\_split()} was applied with a test-set fraction of 33\%. For the classification of 10 materials, an F1 level of approximately 0.97 was obtained. These results represent an important baseline; however, they cannot be considered fully equivalent to the present neural-network experiments, because this work uses a group split by slices and an extended feature representation (\texttt{log1p\_deriv}).

A comparison of the baseline approach with the neural-network models is shown in table~\ref{tab:boosting_comparison}. Because the gradient-boosting and neural-network experiments used different data-splitting protocols and feature representations, their results are not directly comparable. Nevertheless, the final MLP model achieved a macro F1-score comparable to the previously obtained gradient-boosting baseline while being evaluated using the stricter group-splitting protocol.

\begin{table}[htbp]
\centering
\small
\caption{Comparison of the neural-network models with a preliminary gradient-boosting baseline. The results are not directly comparable because different data-splitting protocols and feature representations were used.}
\label{tab:boosting_comparison}
\begin{tabularx}{\textwidth}{p{2.6cm} p{3.0cm} p{2.5cm} X c}
\toprule
Approach & Model & Splitting & Features & Macro F1 / F1 \\
\midrule
Gradient boosting baseline 
& XGBoost / LightGBM / CatBoost 
& row-wise split 
& THL-based ROI features 
& $\sim$0.97 \\

Neural network 
& 1D-CNN 
& group split 
& raw features 
& 0.9461 \\

Neural network 
& MLP 
& group split 
& raw features 
& 0.9649 \\

Neural network 
& MLP 
& group split 
& \texttt{log1p\_deriv} 
& 0.9745 \\
\bottomrule
\end{tabularx}
\end{table}

\subsection{Effect of the \texttt{Empty} class}
\label{subsec:empty}

An additional experiment was performed by including \texttt{Empty} as an eleventh class. In this case, the dataset size increased to 257040 rows. A comparison with the main formulation without \texttt{Empty} is shown in table~\ref{tab:empty}.

\begin{table}[htbp]
\centering
\caption{Effect of including \texttt{Empty} as an additional class.}
\label{tab:empty}
\begin{tabular}{lccc}
\hline
Formulation & Accuracy & Balanced accuracy & Macro F1 \\
\hline
Without \texttt{Empty} & 0.9745 & 0.9745 & 0.9745 \\
With \texttt{Empty} & 0.9283 & 0.9300 & 0.9294 \\
\hline
\end{tabular}
\end{table}

The decrease in performance was mainly associated with confusion between \texttt{Air} and \texttt{Empty}. In the experiment with \texttt{Empty}, the F1-score was 0.80 for \texttt{Air} and 0.80 for \texttt{Empty}. Therefore, \texttt{Empty} was excluded from the main material-classification task and treated as an auxiliary background class, which may be useful for preliminary segmentation or filtering.

\section{Discussion}
\label{sec:discussion}

The obtained results show that ROI spectral features extracted from PCCT data can be used for effective material identification, complementing conventional K-edge and material-decomposition approaches in PCCT~\cite{clark2014spectraldiffusion,Sotenskii_2024}. At the same time, classification performance depends substantially on both the spectral representation and the problem formulation.

The MLP outperformed the 1D-CNN, although the latter may seem a natural choice for ordered spectral data. A possible explanation is that the spectra considered here have relatively low dimensionality, and the informative features may be distributed over the entire THL range rather than be purely local. In this situation, the fully connected architecture can effectively use global relationships between threshold points. This result indicates that, within the considered dataset and acquisition configuration, the absolute reconstructed signal level contains discriminative information.

Large ROIs were classified almost perfectly, as expected, since their spectra are formed from a larger number of pixels and therefore have better counting statistics. Small ROIs represented a more challenging case, because they are more affected by fluctuations due to limited pixel statistics. Nevertheless, the final model trained on the combined ROI dataset achieved high performance, making it more general than a model trained only on large ROIs.

The residual errors of the final model have a physically interpretable origin. The H$_2$O/PMMA pair was the most challenging, since both materials have smooth spectral dependencies without a pronounced K-edge in the considered THL range. Therefore, their separation is based on relatively subtle differences in spectral shape and amplitude, which may be partially masked by statistical fluctuations, especially for small ROIs. The errors between Gd\_1 and Gd\_2 reflect the difficulty of separating two close variants of one sample family. The errors between Nd\_1, La\_1, and Nd\_2 indicate partially overlapping threshold-response patterns. These similarities may be related to the relatively close absorption-edge energies of the lanthanides, detector energy resolution, sample concentration, matrix composition, and reconstruction effects. A direct attribution to K-edge proximity would require detector-threshold calibration in energy units.

The main limitation of the present study is that the dataset was obtained using a single phantom and a single experimental configuration. Although the group split by slice index is a stricter protocol than random row-wise splitting, it still represents internal validation. A stricter assessment of method transferability requires independent scans, other phantoms, other concentrations, other matrices, and possibly other reconstruction parameters.

It is also important to emphasize that the proposed model solves a classification problem between predefined classes from the training dataset. Therefore, its output should not be interpreted as universal identification of Gd, La, or Nd at arbitrary concentration and in an arbitrary matrix. In contrast to algorithms aimed at detecting the K-edge of a specific element and estimating its concentration, the present model recognizes specific spectral patterns represented in the training dataset. Consequently, transferring the method to new concentrations, new matrices, or new acquisition conditions requires expansion of the training dataset and external validation.

The separate analysis of the Gd\_1/Gd\_2 pair shows that the neural-network model can exploit weak but reproducible differences across the full ROI threshold-response curve, rather than relying exclusively on a small number of predefined K-edge-related features. This is an advantage for the classification of specific samples, but it also limits the physical interpretation of the result: the model does not automatically separate the contribution of elemental concentration from the contribution of the matrix and sample-preparation conditions.

\section{Conclusion}
\label{sec:conclusion}

A neural-network-based approach to material identification from ROI spectral features in PCCT data was developed and evaluated. Among the considered models and preprocessing strategies, the MLP combined with the \texttt{log1p\_deriv} representation provided the best performance. For the 10-class task evaluated using a group split by tomographic slice, the final model achieved an accuracy of 0.9745, a balanced accuracy of 0.9745, and a macro F1-score of 0.9745.

Within the considered phantom dataset, the full threshold-dependent ROI responses contained sufficient information to distinguish both different materials and specific samples containing the same target element. However, the trained model classifies predefined experimental sample classes and should not be interpreted as a universal method for determining arbitrary elemental concentrations. Further work should therefore include independent scan series, additional phantoms, varied concentrations and matrices, and external validation under different acquisition and reconstruction conditions.

\section*{Data and code availability}

The processed data and source code supporting the findings of this study are available from the corresponding author upon reasonable request.

\section*{Funding}

This research was funded by the Russian Science Foundation, grant number 22-15-00072-\foreignlanguage{russian}{П}.


\begin{thebibliography}{99}

\bibitem{taguchi2013vision}
K.~Taguchi and J.S.~Iwanczyk,
\emph{Vision 20/20: Single photon counting x-ray detectors in medical imaging},
Med. Phys. \textbf{40} (2013) 100901,
\href{https://doi.org/10.1118/1.4820371}{doi:10.1118/1.4820371}.

\bibitem{llopart2007timepix}
X.~Llopart, R.~Ballabriga, M.~Campbell, L.~Tlustos and W.~Wong,
\emph{Timepix, a 65k programmable pixel readout chip for arrival time, energy and/or photon counting measurements},
Nucl. Instrum. Meth. A \textbf{581} (2007) 485--494,
\href{https://doi.org/10.1016/j.nima.2007.08.079}{doi:10.1016/j.nima.2007.08.079}.

\bibitem{ballabriga2013medipix3rx}
R.~Ballabriga, J.~Alozy, G.~Blaj, M.~Campbell, M.~Fiederle, E.~Frojdh,
E.H.M.~Heijne, X.~Llopart, M.~Pichotka, S.~Procz, L.~Tlustos and W.~Wong,
\emph{The Medipix3RX: a high resolution, zero dead-time pixel detector readout chip allowing spectroscopic imaging},
JINST \textbf{8} (2013) C02016,
\href{https://doi.org/10.1088/1748-0221/8/02/C02016}{doi:10.1088/1748-0221/8/02/C02016}.

\bibitem{dudak2020timepix}
J.~Dudak,
\emph{High-resolution X-ray imaging applications of hybrid-pixel photon counting detectors Timepix},
Radiat. Meas. \textbf{137} (2020) 106409,
\href{https://doi.org/10.1016/j.radmeas.2020.106409}{doi:10.1016/j.radmeas.2020.106409}.

\bibitem{procz2013medipix3ct}
S.~Procz, K.-A.~Wartig, A.~Fauler, A.~Zwerger and M.~Fiederle,
\emph{Medipix3 CT for material sciences},
JINST \textbf{8} (2013) C01025,
\href{https://doi.org/10.1088/1748-0221/8/01/C01025}{doi:10.1088/1748-0221/8/01/C01025}.

\bibitem{suslova2022nanomaterials}
E.V.~Suslova, A.P.~Kozlov, D.A.~Shashurin, V.A.~Rozhkov, R.V.~Sotenskii,
S.V.~Maximov, S.V.~Savilov, O.S.~Medvedev and G.A.~Chelkov,
\emph{New Composite Contrast Agents Based on Ln and Graphene Matrix for Multi-Energy Computed Tomography},
Nanomaterials \textbf{12} (2022) 4110,
\href{https://doi.org/10.3390/nano12234110}{doi:10.3390/nano12234110}.

\bibitem{suslova2022lafml}
E.~Suslova, D.~Shashurin, A.~Kozlov, S.~Maximov, V.~Rozhkov, R.~Sotenskii,
S.~Savilov, O.~Medvedev and G.~Chelkov,
\emph{Development of La--graphene composite contrasting agents for photon-counting computed tomography},
Functional Materials Letters \textbf{15} (2022) 2250029,
\href{https://doi.org/10.1142/S1793604722500291}{doi:10.1142/S1793604722500291}.

\bibitem{schlomka2008kedge}
J.-P.~Schlomka, E.~Roessl, R.~Dorscheid, S.~Dill, G.~Martens, T.~Istel,
C.~B\"aumer, C.~Herrmann, R.~Steadman, G.~Zeitler, A.~Livne and R.~Proksa,
\emph{Experimental feasibility of multi-energy photon-counting K-edge imaging in pre-clinical computed tomography},
Phys. Med. Biol. \textbf{53} (2008) 4031--4047,
\href{https://doi.org/10.1088/0031-9155/53/15/002}{doi:10.1088/0031-9155/53/15/002}.

\bibitem{simohamed2021atherosclerosis}
S.A.~Si-Mohamed, M.~Sigovan, J.C.~Hsu, V.~Tatard-Leitman,
L.~Chalabreysse, P.C.~Naha et al.,
\emph{In vivo molecular K-edge imaging of atherosclerotic plaque using photon-counting CT},
Radiology \textbf{300} (2021) 98--107,
\href{https://doi.org/10.1148/radiol.2021203968}
{doi:10.1148/radiol.2021203968}.

\bibitem{boccalini2023gadolinium}
S.~Boccalini, R.~Dessouky, P.-A.~Rodesch, H.~Lacombe,
Y.~Yagil, E.~Lahoud et al.,
\emph{Gadolinium K-edge angiography with a spectral photon-counting CT in atherosclerotic rabbits},
Diagn. Interv. Imaging \textbf{104} (2023) 490--499,
\href{https://doi.org/10.1016/j.diii.2023.05.002}
{doi:10.1016/j.diii.2023.05.002}.

\bibitem{jost2023kedge}
G.~Jost, M.~McDermott, R.~Gutjahr, T.~Nowak, B.~Schmidt and H.~Pietsch,
\emph{New contrast media for K-edge imaging with photon-counting detector CT},
Invest. Radiol. \textbf{58} (2023) 515--522,
\href{https://doi.org/10.1097/RLI.0000000000000978}
{doi:10.1097/RLI.0000000000000978}.

\bibitem{clark2014spectraldiffusion}
D.P.~Clark and C.T.~Badea,
\emph{Spectral diffusion: an algorithm for robust material decomposition of spectral CT data},
Phys. Med. Biol. \textbf{59} (2014) 6445--6466,
\href{https://doi.org/10.1088/0031-9155/59/21/6445}{doi:10.1088/0031-9155/59/21/6445}.

\bibitem{Sotenskii_2024}
R.V.~Sotenskii, V.A.~Rozhkov, D.A.~Shashurin, E.V.~Suslova and G.A.~Chelkov,
\emph{Novel algorithm for qualitative and quantitative material analysis by the K-edges for photon-counting computed tomography},
JINST \textbf{19} (2024) P04009,
\href{https://doi.org/10.1088/1748-0221/19/04/P04009}{doi:10.1088/1748-0221/19/04/P04009}.

\bibitem{abascal2021material}
J.F.P.J.~Abascal, N.~Ducros, V.~Pronina, S.~Rit, P.-A.~Rodesch, T.~Broussaud,
S.~Bussod, P.~Douek, A.~Hauptmann, S.~Arridge and F.~Peyrin,
\emph{Material decomposition in spectral CT using deep learning: a Sim2Real transfer approach},
IEEE Access \textbf{9} (2021) 25632--25647,
\href{https://doi.org/10.1109/ACCESS.2021.3056150}{doi:10.1109/ACCESS.2021.3056150}.

\bibitem{bousse2024review}
A.~Bousse, V.S.S.~Kandarpa, S.~Rit, A.~Perelli, M.~Li, G.~Wang, J.~Zhou and G.~Wang,
\emph{Systematic review on learning-based spectral CT},
IEEE Trans. Radiat. Plasma Med. Sci. \textbf{8} (2024) 113--137,
\href{https://doi.org/10.1109/TRPMS.2023.3314131}{doi:10.1109/TRPMS.2023.3314131}.

\bibitem{rajagopal2025multimaterial}
J.R.~Rajagopal, S.~Rapaka, F.~Farhadi et al.,
\emph{Development of a deep learning based approach for multi-material decomposition in spectral CT: a proof of principle in silico study},
Sci. Rep. \textbf{15} (2025) 28814,
\href{https://doi.org/10.1038/s41598-025-09739-9}{doi:10.1038/s41598-025-09739-9}.

\bibitem{sourceray_sb120350}
Source-Ray, Inc.,
\emph{SB-120 X-ray generators: SB-120-350},
\href{https://sourceray.com/x-ray-generators/products/family/sb-120}
{Source-Ray product information},
accessed 14 July 2026.

\bibitem{jakubek2014widepix}
J.~Jakubek, M.~Jakubek, M.~Platkevic, P.~Soukup, D.~Turecek,
V.~Sykora and D.~Vavrik,
\emph{Large area pixel detector WIDEPIX with full area sensitivity composed of 100 Timepix assemblies with edgeless sensors},
JINST \textbf{9} (2014) C04018,
\href{https://doi.org/10.1088/1748-0221/9/04/C04018}
{doi:10.1088/1748-0221/9/04/C04018}.

\bibitem{vanAarle2016astra}
W.~van Aarle, W.J.~Palenstijn, J.~Cant, E.~Janssens, F.~Bleichrodt,
A.~Dabravolski, J.~De Beenhouwer, K.J.~Batenburg and J.~Sijbers,
\emph{Fast and flexible X-ray tomography using the ASTRA Toolbox},
Opt. Express \textbf{24} (2016) 25129--25147,
\href{https://doi.org/10.1364/OE.24.025129}
{doi:10.1364/OE.24.025129}.

\bibitem{paszke2019pytorch}
A.~Paszke et al.,
\emph{PyTorch: An imperative style, high-performance deep learning library},
in \emph{Advances in Neural Information Processing Systems 32
(NeurIPS 2019)}, article 721 (2019).

\bibitem{pedregosa2011scikit}
F.~Pedregosa et al.,
\emph{Scikit-learn: Machine Learning in Python},
J. Mach. Learn. Res. \textbf{12} (2011) 2825--2830.

\bibitem{loshchilov2019adamw}
I.~Loshchilov and F.~Hutter,
\emph{Decoupled Weight Decay Regularization},
in \emph{Proc. International Conference on Learning Representations}
(ICLR 2019),
arXiv:1711.05101,
\href{https://doi.org/10.48550/arXiv.1711.05101}
{doi:10.48550/arXiv.1711.05101}.

\end{thebibliography}
\end{document}